# Terminal polydispersity in the crystallization of polydisperse Lennard-Jones liquid

Sarmistha Sarkar, Mantu Santra and Biman Bagchi*

***Solid State and Structural Chemistry Unit,***

***Indian Institute of Science, Bangalore 560012, India.***

***E-mail:** bbagchi@sscu.iisc.ernet.in*

**We find through computer simulations that the fractional volume change on freezing of polydisperse Lennard-Jones liquid decreases with increasing polydispersity and approaches zero near a *terminal polydispersity* of 0.11, independent of temperature. The transition however remains *first order* at terminal polydispersity. Average inherent structure (IS) energy of the crystalline phase increases nearly quadratically with polydispersity indices (δ) and marked by a *crossover* to nearly constant IS energy in the amorphous phase.**

## I. Introduction

Polydisperse colloidal systems are known to be good glass formers. Many studies suggest that beyond a transition polydispersity, crystallization never takes place and the amorphous phase becomes the true ground state of the system [**1-4**]. This transition polydispersity and the nature of transition at this point are subjects of great interest. The absence of liquid-solid transition in polydisperse systems has been attributed to the increase in surface tension, leading to the difficulty of nucleating the crystalline phase [**3**]. A density functional theory (DFT) analysis of freezing in hard sphere fluid finds a terminal polydispersity of 0.048, followed by a re-entrant melting at large density [**4**]. Experimental studies, however, observe a nearly universal value 0.12 for the terminal polydispersity [**5**]. Simulation studies on hard sphere crystals also predict a possible terminal polydispersity of $0.12 \pm 0.01$ [**6**] whereas simple mean-field model of polydisperse hard spheres suggests the value to be 0.0833 [**7**]. It is well-known that for binary systems, Hume-Rothery rule precludes freezing when the radii of the two components differ by more than 15%, beyond which glass becomes the stable phase at high density [**8**].

As polydispersity increases, the entropy of the fluid also increases. Thus, it is expected that freezing will occur at higher and higher volume fraction φ as the polydispersity index δ increases (both quantified below). In hard sphere system, this increase in entropy loss on freezing must be counter-balanced by an increase (in absolute magnitude) of the PΔV term, where P is the pressure and ΔV is the volume change on freezing. At terminal polydispersity both the PΔV term and entropy loss become zero. For Lennard-Jones system, an important contribution is made by the change in the internal energy, ΔE term. Polydispersity dependence of this energy term makes freezing of L-J liquid quite different from that of the hard sphere system, as we report here.

While a sharp increase in surface tension with polydispersity can exclude formation of the crystalline nuclei [**3**], it still leaves open the question of the existence of a stable crystalline phase, at a large polydispersity, marked by a global free energy minimum. DFT analysis, on the other hand, suggests that the crystalline minimum either disappears or becomes metastable with respect to the minimum corresponding to the amorphous phase beyond terminal polydispersity. The situation is thus a bit like spinodal decomposition. In the DFT analysis, the reduced stability of the crystalline phase arises from the reduced value of the first peak of the static structure factor, S(k). This is then related to the larger disorder in polydisperse fluids than monodisperse fluids and this disorder increases with polydispersity. However, the said DFT analysis, in addition to predict a lower value of the terminal polydispersity for hard spheres, does not address the issue of the transition properties, like the polydispersity dependence of the volume fraction change across the transition [**4**]. Temperature dependence of the transition has not been studied theoretically (most studies considered only hard spheres) and thus the phase diagram has not been fully elucidated yet.

In this article we present the remarkable result that the transition polydispersity *converges to a maximum value of 0.11, termed as terminal polydispersity, at the large volume fraction* ( in agreement with experiments) when plotted against volume fraction at several temperatures. The fractional volume change on freezing *approaches zero* as the polydispersity is increased at constant temperature, but the orientational order parameter $Q_6$ (defined below) shows a large amplitude jump as the system passes through the transition polydispersity. In addition, for a particular temperature, compressibility is found to decrease till the transition polydispersity and this we show by inherent structure analysis [**9**] that some of the above results can be understood from the increase in the internal energy of the crystalline phase with polydispersity.

## II. Simulation details

We carry out MD simulation of LJ polydisperse particles for 50,000 steps in NVT ensemble to fix constant temperature followed by 50,000 steps in NVE ensemble to equilibrate. After equilibration 500000 more steps have been carried out in NVE ensemble for acquiring results. The system is composed of 500 particles with periodic boundary condition and with polydispersity in size. Polydispersity in size is described by a Gaussian distribution of particle diameters $\sigma$, $P(\sigma)=\frac{1}{\sqrt{2\pi d^2}}\exp\left[-\frac{1}{2}\left(\frac{\sigma-\bar{\sigma}}{d}\right)^2\right]$, that also defines the polydispersity index $\delta=\frac{d}{\bar{\sigma}}$ where d is standard deviation of the distribution and $\bar{\sigma}$ is the mean diameter. We employ simulations for different polydispersity indices S = 0 to 0.20 in steps of 0.01. The particles interact with Lennard-Jones potential. The depth of the potential (ε) is assumed to have the same value for all particle pairs. The potential for a pair of particles *i* and *j* is cut and shifted to zero at the distance $r_c = 2.5\sigma_{ij}$, where $\sigma_{ij} = (\sigma_i + \sigma_j)/2$. We employ NVT and NVE ensembles in both MD and Monte Carlo computer simulations to obtain the phase diagram, and use the standard conjugate gradient method to obtain the inherent structures along an MD trajectory. We carry out NPT simulation to calculate fractional volume change and compressibility.

## III. Results and discussions

### A. Liquid-Solid phase diagram

**Fig. 1(a)** depicts the liquid-solid phase diagram of the system. Here we plot the transition polydispersity versus volume fraction at three different temperatures. The plot shows that with the increase in volume fraction φ, there is initial an increase in transition polydispersity and then it becomes invariant of φ. The value of the transition polydispersity converges to a maximum value of $\delta_t \sim 0.11$, termed as terminal polydispersity. The significance of the plot is that for polydispersity higher than $\delta_t$, there is no crystallization and the system inhabits glassy minima. We find that the terminal polydispersity ranges between the narrow range bounded by 0.09 and 0.11 – the lower limit is obtained at high temperature.

As already mentioned, the terminal polydispersity observed in experiments is close to 0.12 in colloidal spheres which may be modeled as hard spheres [**5**]. In LJ fluids, the *transition polydispersity* is dependent on temperature and it converges to an almost constant value of 0.11 (terminal polydispersity), for all temperatures as shown in **Fig. 1(a).** We found no signature of re-entrant melting in our polydisperse LJ liquid. We checked that both by Monte Carlo, molecular dynamics and also free energy calculations.

### B. The fractional volume change on freezing

**Figure 1(b)** shows the remarkable result that with the increase in polydispersity, the fractional volume change on freezing decreases, and eventually becomes very small near the terminal polydispersity. The freezing transition remains first order as discussed below. Results of Figure 1(b) can be understood from decrease of compressibility (at the spinodal/transition point) with δ at the transition density as shown in **Figure 1(c)**.

### C. Calculation of compressibility

**Figure 1(c)** shows that with the increase in polydispersity, the value of compressibility $(\chi)$ (at the spinodal/transition point) decreases gradually till transition polydispersity 0.09 and then $\chi$ remains almost constant with the increase in polydispersity. This signifies the structural change at $\delta = 0.09$ where the system enters the amorphous phase.

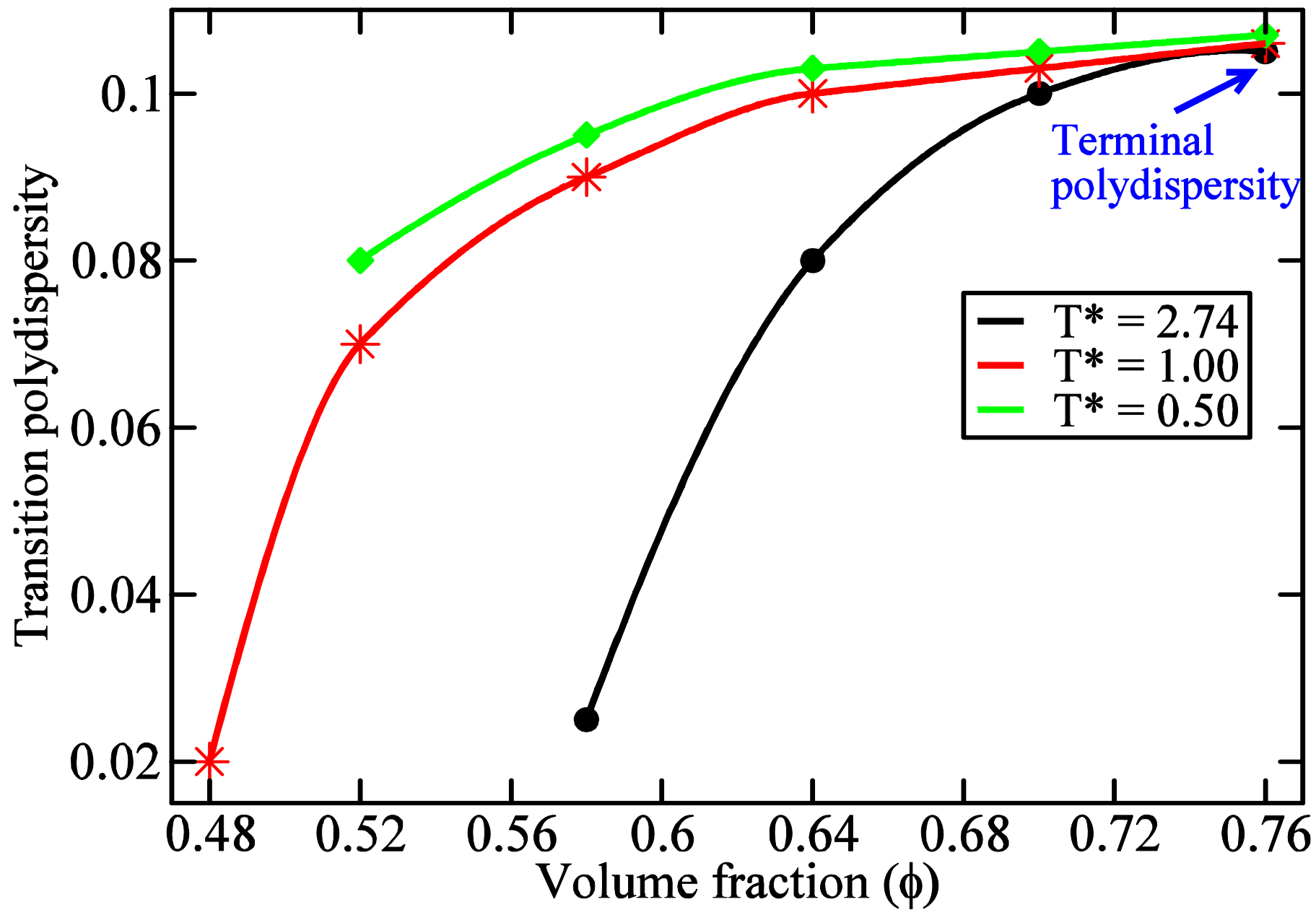

Fig. 1(a)

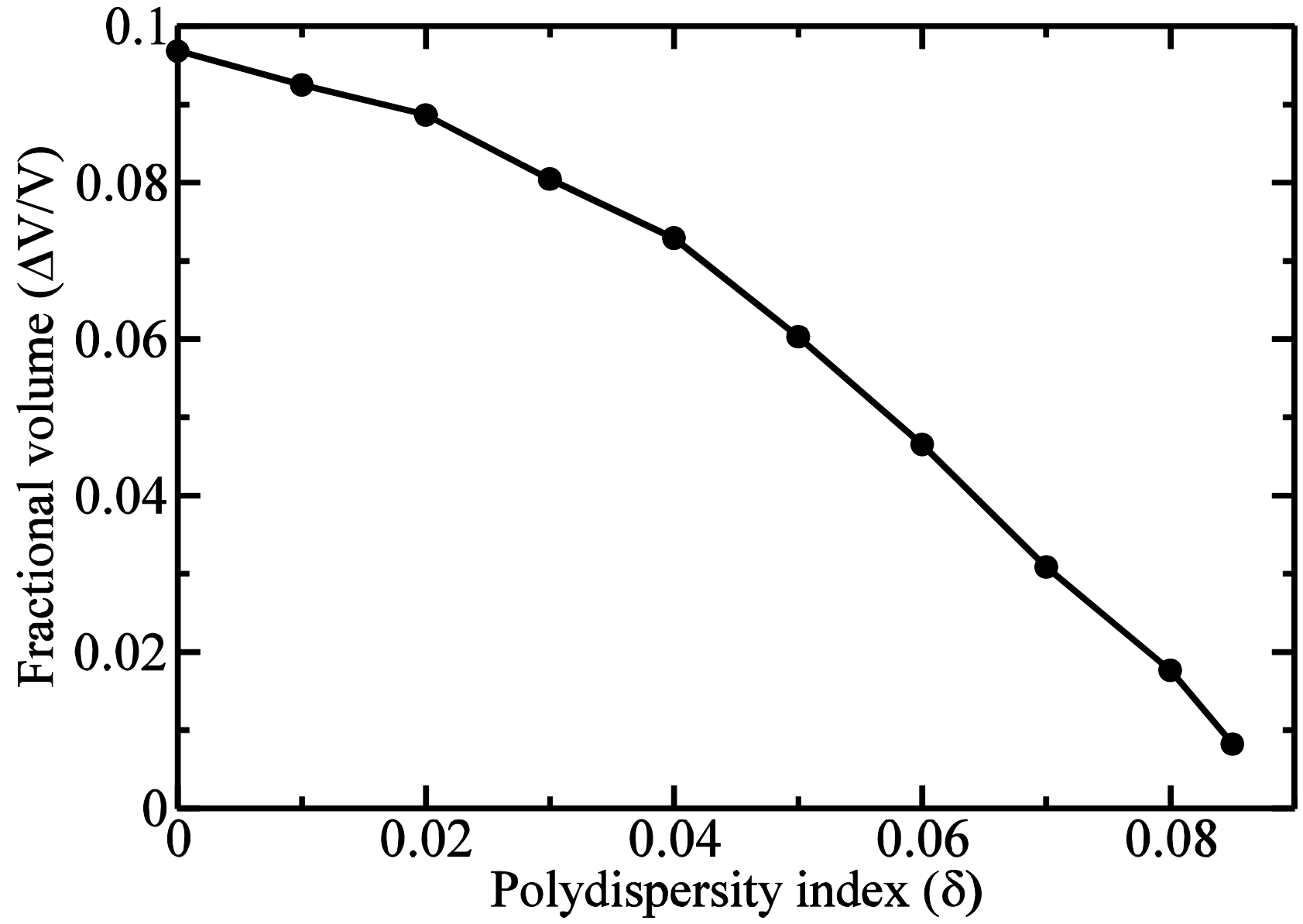

Fig. 1(b)

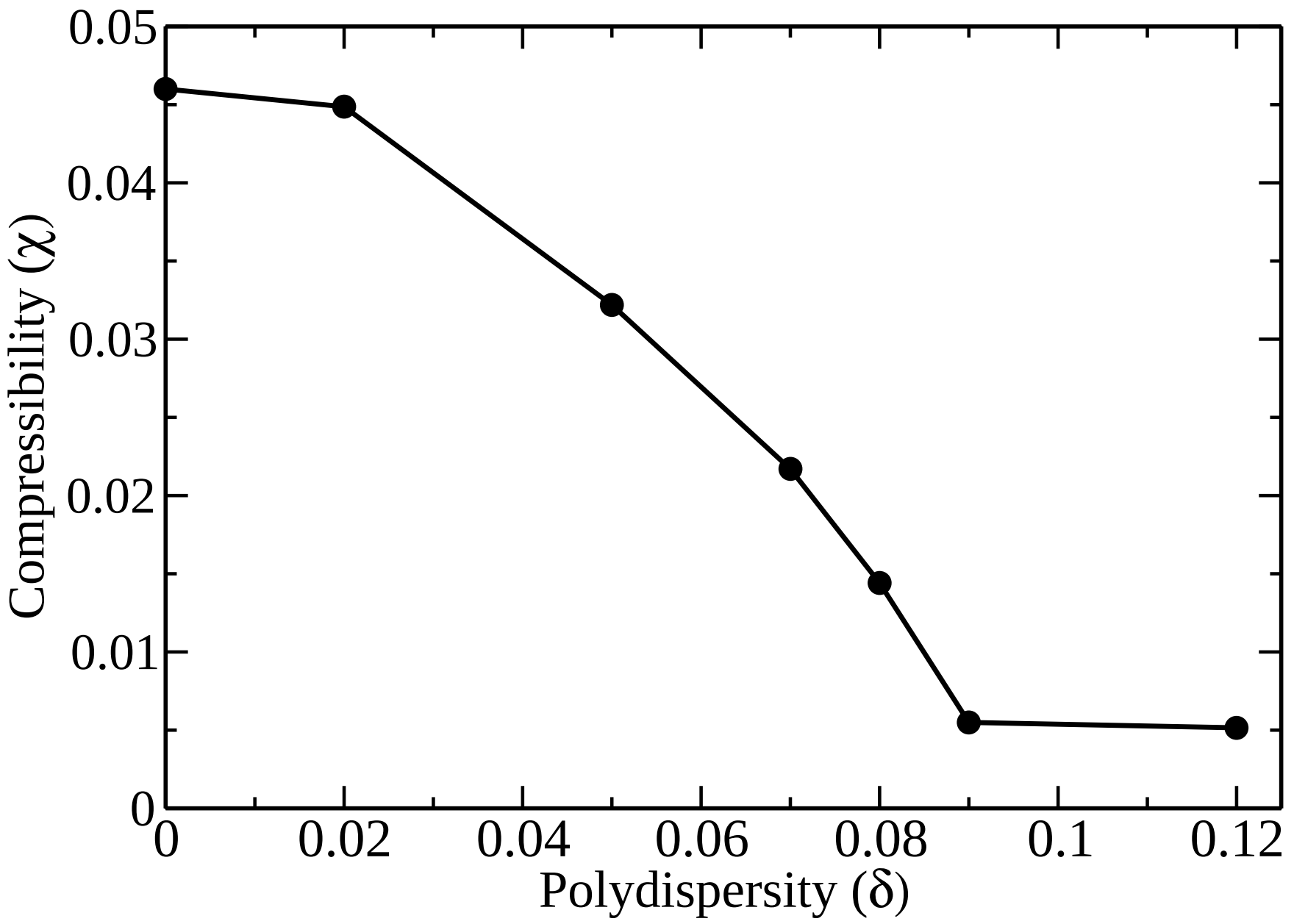

Fig. 1(c)

**Fig. 1** **(a).** Phase diagram (transition polydispersity versus volume fraction) at three different temperatures T* = 0.50, 1.0 and 2.74. The values of the transition polydispersity at different temperatures converge to a maximum value of $\delta_t \sim 0.11$, termed as terminal polydispersity. **(b).** Plot of variation of fractional volume change with polydispersity index δ. Note that the fractional volume change on freezing approaches zero as the polydispersity is increased at constant temperature T* =1.0. **(c).** Plot of variation of compressibility at transitin point with polydispersity index $\delta$. Note that at constant temperature T* =1.0, as the polydispersity is increased, the value of compressibility decreases and then becomes almost constant from $\delta = 0.09$ onwards.

# IV. Quantification of structural change

## A. Computation of orientational order

To quantify the variation in the amplitude of structural change on freezing with polydispersity, we compute orientational ($Q_6$) and translational order (τ). Following Steinhardt *et al.* [**10**] a vector $r_{ij}$ pointing from a given molecule (i) to one of its nearest neighbors (j) is denoted as a "bond". For each bond one determines the quantity, $Q_{lm}(r_{ij}) = Y_{lm}(\theta_{ij}, \phi_{ij})$, where $\hat{r}_{ij}$ is the unit vector of $r_{ij}$ with the related polar and azimuthal angles $\theta_{ij}$ and $\phi_{ij}$ and $Y_{lm}$ is the associated spherical harmonics. An average over all bonds is performed to obtain, $\overline{Q_{lm}} = < Q_{lm}(\hat{r}_{ij}) >$. The rotationally invariant order metrics (i.e. the measure of orientational order) is computed as, $Q_l = \left[ \frac{4\pi}{2l+1} \sum_{m=-1}^{m=1} \overline{Q_{lm}}^2 \right]^{1/2}$. In fcc lattice the orientational order is characterized by six fold symmetry that corresponds to l = 6. Fig. 2(a) shows a sharp structural transition from fcc solid to fluid at 9% polydispersity indicated by a jump in $Q_6$ at T = 1.0, φ = 0.58, confirming the first order nature of the phase transition, even at $\delta = 0.09$.

B. **Computation of translational order**

We also compute the translational order (τ) from the radial distribution function [**11**] as, $\tau = \frac{1}{s_c} \int_0^{s_c} \left[ g(s) - 1 \right] ds$, where $s = r\rho^{\frac{1}{3}}$ is the radial distance scaled by the number density, *g*(*s*) is the pair correlation function, and $s_c$ is the upper limit of *s* and set to 3.5. Translational order provides a measure of the local density modulations over a finite number of coordination shells and can be considered as a quantity to characterize the local structure. In Fig. 2(b) the translational order also shows a transition similar to $Q_6$ at the same polydispersity. Both Fig. 2(a) and 2(b) together confirm a structural transition at 9% polydispersity at T = 1.0, φ = 0.58.

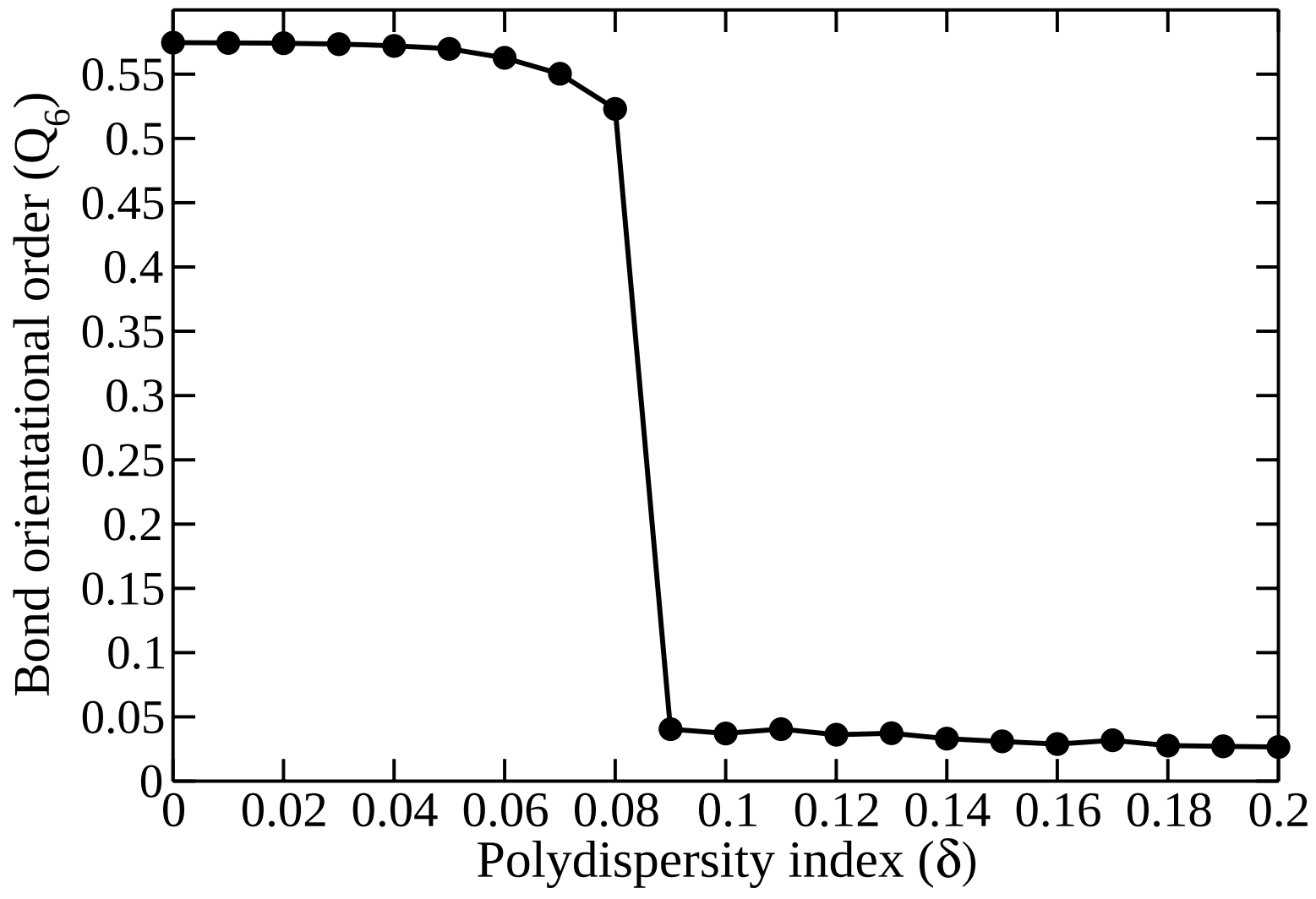

Fig. 2 (a)

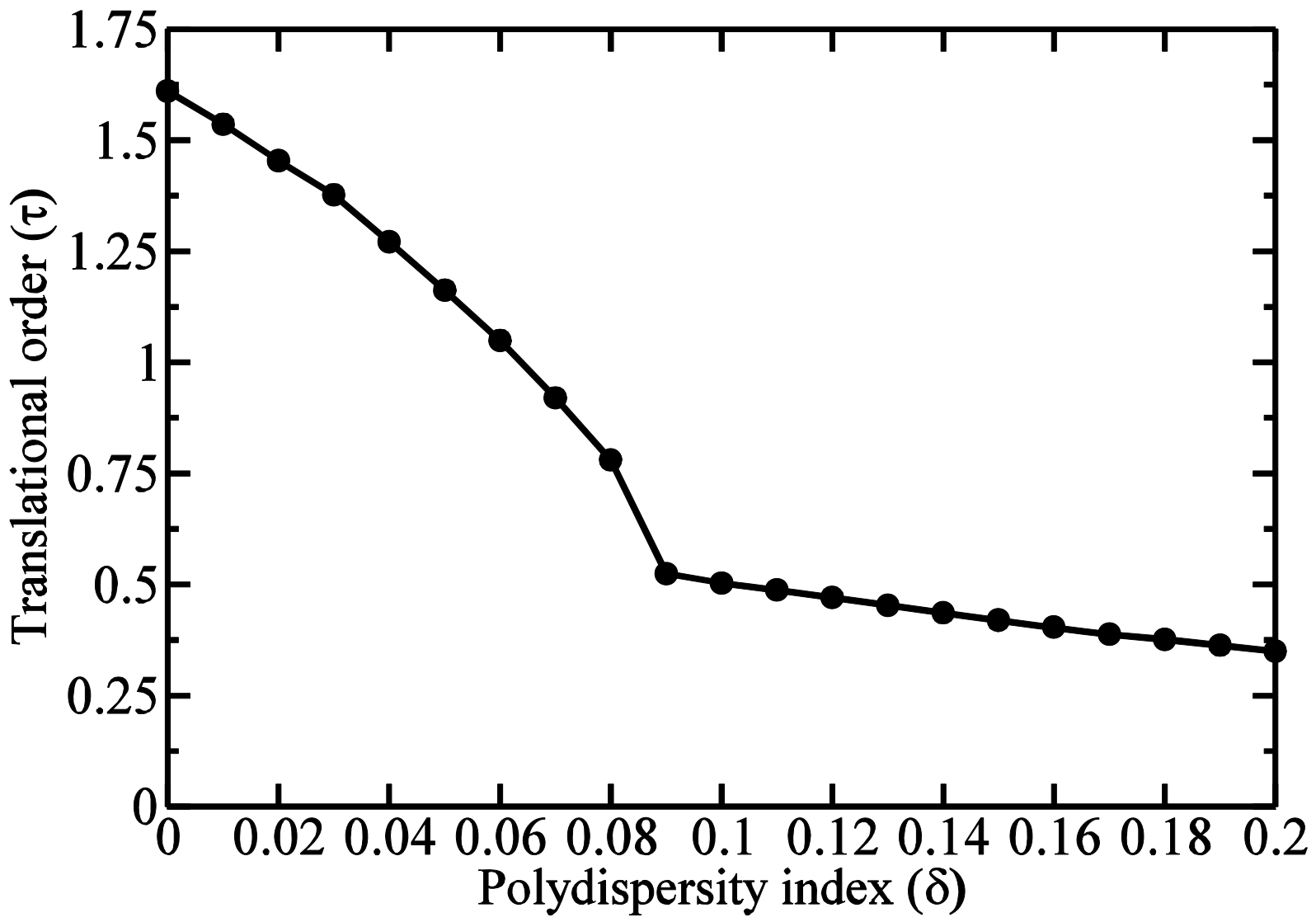

Fig. 2 (b)

**Fig. 2**. Plot of (a) bond orientational order parameter and (b) translational order parameter with respect to polydispersity index δ. Note that the plots of translational and orientational order together confirm a structural change at 9% polydispersity at temperature T = 1.0, volume fraction φ = 0.58.

## V. Insight into freezing by studying inherent structure of liquid

### A. Calculation of average inherent structure energy

Polydispersity makes first principle study of the freezing transition a bit difficult because analytical solution of pair correlation functions is hard to obtain. Fortunately, one can obtain considerable insight into the freezing process by studying the inherent structures of the liquids and solids and also their energy as a function of the polydispersity. **Fig. 3(a)** shows the variation of average inherent structure energy (*AISE)* against polydispersity index. *AISE* increases with polydispersity till a critical value of around $\delta = 0.09$ (at $T^* = 1.0$, $\varphi = 0.58$) and then it becomes almost invariant of $\delta$ which indicates a structural change at the polydispersity $\delta = 0.09$.

### B. Distribution of average inherent structure energy

**Figure** 3(b) shows distribution of IS energy at different polydispersity indices. Note the lack of spread in IS energy till $\delta = 0.085$ (at $T^* = 1.0$, and $\varphi = 0.58$). There is a complete change in distribution from polydispersity 0.09 and larger.

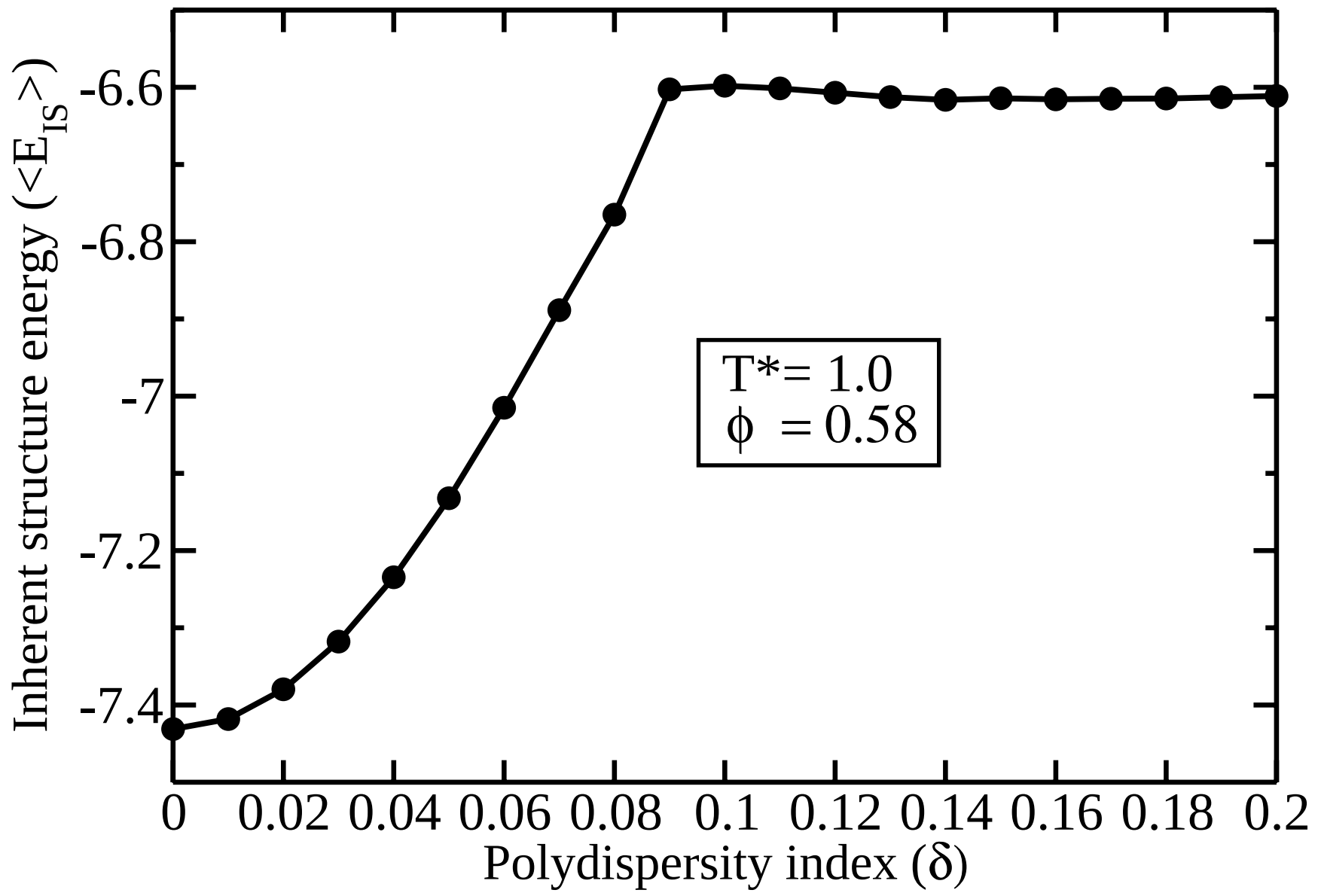

Inherent structure energy (<E_IS>)
Polydispersity index (δ)
T*= 1.0
ϕ = 0.58
-6.6
-6.8
-7
-7.2
-7.4
0 0.02 0.04 0.06 0.08 0.1 0.12 0.14 0.16 0.18 0.2

Fig. 3(a)

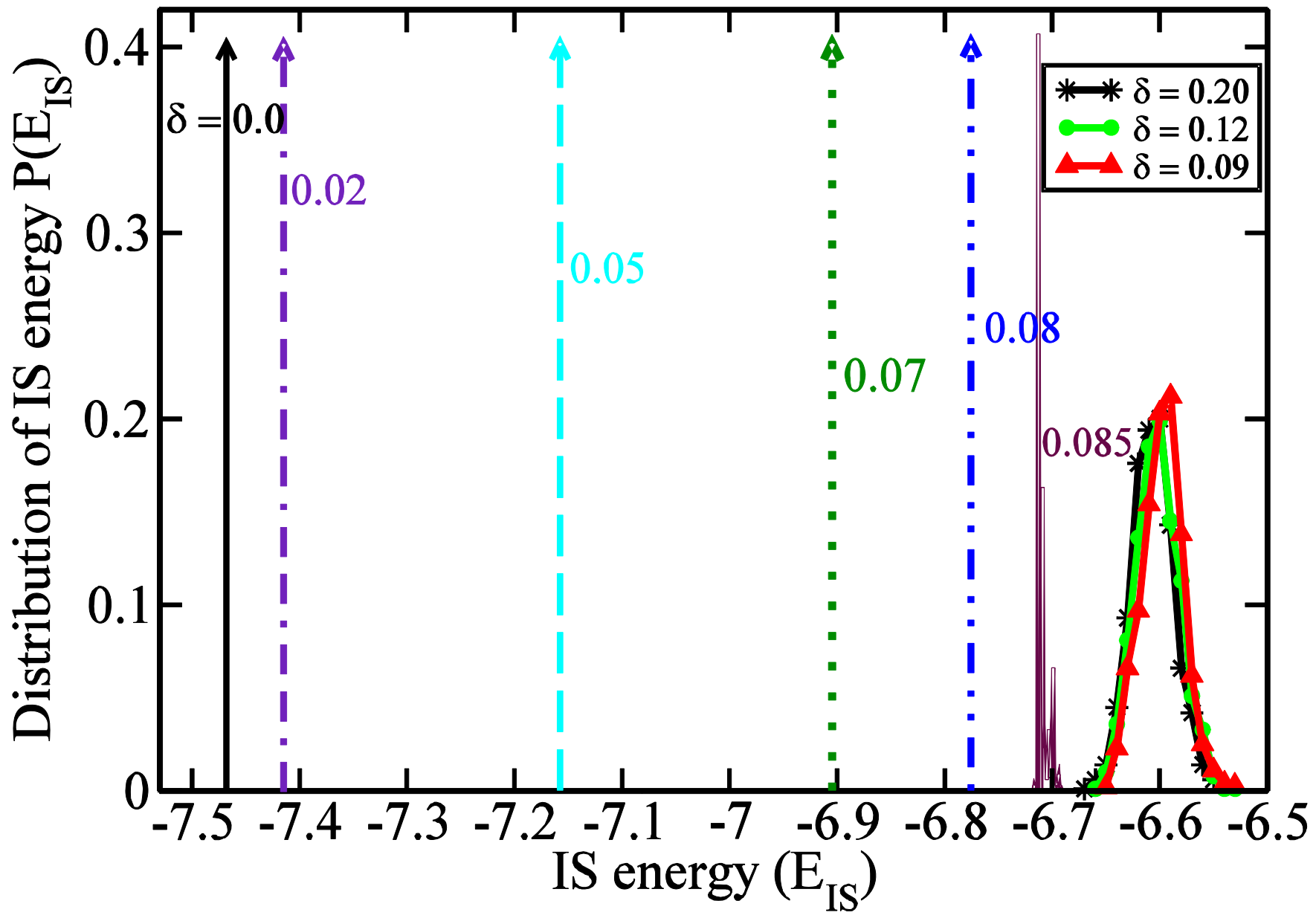

Distribution of IS energy P(E_IS)
IS energy (E_IS)
δ = 0.0
0.02
0.05
0.07
0.08
0.085
δ = 0.20
δ = 0.12
δ = 0.09
0.4
0.3
0.2
0.1
0
-7.5 -7.4 -7.3 -7.2 -7.1 -7 -6.9 -6.8 -6.7 -6.6 -6.5

Fig. 3(b)

Fig. 3(a). Variation of average inherent structure energy with respect to polydispersity index at constant volume fraction 0.58 and temperature T* =1.0 (b) Distribution of IS energy at different polydispersity indices. Note that there is a sharp change in the distribution beyond $\delta = 0.09$.

## IV. Summary and conclusions

The freezing of polydisperse fluids presents an interesting case because the energy-entropy balance becomes increasingly unfavorable for the solid as the polydispersity increases. The energy of the solid increases due to build up of strain energy while the entropy of the liquid increases. These two factors lead to the existence of a terminal polydispersity. What is the state of the system then beyond the terminal polydispersity? The system at large volume fraction and large polydispersity remains in an amorphous state, characterized by multiple configurations and a broad two particle radial distribution function. These results suggest that the amorphous state becomes the global minimum of the free energy surface at large polydispersity.

In an earlier paper, Rice et al. suggested *that the spinodal point of liquid-solid transition of a Lennard-Jones fluid is effectively a random close packing state of hard spheres* and that *no freezing transition is possible beyond this point* [**12**]. It is interesting to find correlation of the present study with this earlier work. This is also in agreement with simulation results of **Ref.6**.

The phase diagram (pressure-temperature or temperature-density) of the system poses interesting challenge. In the case of uniform polydispersity, Gibbs' Phase Rule should be modified to $F = C - P + 3$ [**13**]. The three thermodynamic parameters to be fixed are pressure, temperature and polydispersity.

**Acknowledgement.** We thank Dr. B Jana and Prof. C. Dasgupta for many helpful discussions. This work was supported in parts by grants from DST and CSIR (India). BB thanks DST for a JC Bose Fellowship.

---------------------------------